\documentclass[aps,pre,superscriptaddress,twocolumn]{revtex4-1}
\usepackage{bm}
\usepackage{graphicx}
\usepackage{tikz}
\usepackage{amsmath}
\usepackage{float}
\usepackage{amsfonts}
\usepackage{graphicx}
\usepackage{xcolor}
\usepackage{epstopdf}
\usepackage{booktabs}

\begin{document}

\title{Selection and control of pathways by using externally adjustable noise on a stochastic cubic autocatalytic chemical system}

\author{Jean-S\'{e}bastien Gagnon}
\email[]{gagnon01@fas.harvard.edu}
\affiliation{Department of Earth and Planetary Sciences, Harvard University, Cambridge, Massachusetts, USA}
\affiliation{Natural Sciences Department, Castleton University, Castleton, Vermont, USA}

\author{David Hochberg}
\email[]{hochbergd@cab.inta-csic.es}
\affiliation{Department of Molecular Evolution, Centro de
Astrobiolog\'{i}a (CSIC-INTA), Torrej\'{o}n de Ard\'{o}z, Madrid,
Spain}

\author{Juan P\'{e}rez-Mercader}
\email[]{jperezmercader@fas.harvard.edu}
\affiliation{Department of Earth and Planetary Sciences, Harvard University, Cambridge, Massachusetts, USA}
\affiliation{Santa Fe Institute, Santa Fe, New Mexico, USA}

\begin{abstract}
We investigate the effect of noisy feed rates on the behavior of a cubic autocatalytic chemical reaction model. By combining the renormalization group and stoichiometric network analysis, we demonstrate how externally adjustable random perturbations (extrinsic noise) can be used to select reaction pathways and therefore control reaction yields.  This method is general and provides the means to explore the impact that changing statistical parameters in a noisy external environment (such as noisy feed rates, fluctuating reaction rates induced by noisy light, etc) has on chemical fluxes and pathways, thus demonstrating how external noise may be used to control, promote, direct and optimize chemical progress through a given reaction pathway.
\end{abstract}

\date{\today}

\maketitle

\section{Introduction}
\label{sec:Introduction}

It is well-known that fluctuations and noise can affect the behavior
of various systems, from running coupling constants in high-energy
particle physics (e.g.~\cite{Peskin_Schroeder_1995}) to phase
transitions in condensed matter physics
(e.g.~\cite{ZinnJustin_2002}) and noise-induced transitions in
complex
systems~(e.g.~\cite{Garcia_Sancho_1999,Horsthemke_Lefever_2006}). In
particular, noise can affect the dynamics of chemical reactions. For
example, it has been shown that external mechanical noise (shaking vs. stirring) changes the output of chemical replicator reactions~\cite{Carnall_etal_2010}  and that
coherence resonance can be induced in the Belousov-Zhabotinsky (BZ)
reaction by external colored
noise~\cite{Simakov_PerezMercader_2013,Simakov_PerezMercader_2013b}.

Systems undergoing chemical reactions offer a promising and fertile field where selection effects due to external noise can be tested, observed and refined for specific purposes in mind.  The realm of complex chemical phenomena that could be manipulated and controlled in this way with external noise includes sustained chemical oscillations~\cite{Hou_Xin_1999,Simakov_PerezMercader_2013,Simakov_PerezMercader_2013b,Srivastava_etal_2018}, pattern formation~\cite{Lesmes_etal_2003,Biancalani_etal_2016},  excitable dynamics and front propagation~\cite{Garcia_Schimansky_1999,Sendi_etal_2000,Zhou_etal_2002,Lindner_etal_2004}, or any non-linear chemical systems where the number of constituents is sufficiently large so as to allow smooth concentrations to be defined~\footnote{For systems with low number of constituents, both external noise (e.g. temperature, luminosity, stirring, etc) and internal noise (quantum mechanical randomness, may affect the dynamic of the chemical system.  The treatment of internal noise requires more refined tools than the one used in this paper, as discussed in the review~\cite{Tauber_etal_2005}.  See also Ref.~\cite{Hochberg_etal_2005,Cooper_etal_2014} for a study of internal noise in the CAR model.}.

The purpose of this paper is to demonstrate how external experimentally adjustable noise can be used to control and \textit{select} chemical pathways~\footnote{Note that due to the external energy provided by the noise, this selection of chemical pathways may include pathways previously inaccessible to the system.  Although this behavior is not present in the simple cubic autocatalytic reaction used to illustrate our method, there is no theoretical constraint that precludes this in a more complex chemical system.}.  By adjustable noise we mean any external condition on a chemical system (feed rate, illumination, etc) that can be varied in a stochastic manner, and for which the statistical properties (amplitude, spectral exponent, etc) can be varied experimentally. Applying light to the photosensitive dioxide-iodine-malonic acid reaction in order to induce the disappearance of Turing structures is only one an example of the above~\cite{Horvath_etal_1999}.   These stochastic variations in external conditions induce fluctuations in the chemical concentrations, which manifest as noise-dependent modifications of the chemical kinetics.  We illustrate our method using a cubic autocatalytic reaction subjected to a noisy feed rate with gaussian power-law statistics, but the method developed here is general and can be applied to other chemical systems subjected to other types of noise.

To demonstrate how chemical pathways can be selected using noise, we combine
stoichiometric network analysis (SNA) with the dynamic
renormalization group (RG). SNA
is a powerful algebraic method used to study  both the dynamics and
stationarity properties of chemical reactions~\cite{Clarke80,Clarke88,Clarke81}. The pathway
architecture and topology of any chemical reaction network can be
elucidated using SNA. The method is based on convex analysis
\cite{Rockafeller}, and determines a unique set of extreme currents
or extreme flux modes (EFM) which correspond to the edges of a
convex polyhedral flux cone in a Euclidean
reaction-rate space. Following this algebraic technique, all
possible stationary fluxes are then represented by positive linear
combinations of these cone edge vectors. SNA furnishes an efficient method for determining the stability
of non-equilibrium steady states by focusing on the behavior of
steady-state reaction rates, their associated matter fluxes, chemical pathways, and the extreme currents involving the major subnetworks of the overall chemical mechanism.

The renormalization group allows us to compute and express the effect of
fluctuations operating at shorter or longer scales on a system as a non-trivial rescaling of the
parameters of the system.  For chemical reactions, those parameters
are typically the decay rates ($r$) and reaction rates ($\lambda$). In the context of SNA,
the important parameters characterizing the dynamics of the reaction
(i.e. which chemical pathway is predominant) are the inverse
stationary concentrations ($h$) and the convex parameters ($j$).
These latter parameters represent the strength of the matter-fluxes
traversing a specific chemical pathway. Noise affects both ($h,j$),
and the renormalization group allows us to compute their scaling
(or ``running'') as a function of the properties of the
noise~\cite{Hochberg_etal_2003,Zorzano_etal_2004,Gagnon_PerezMercader_2017,Gagnon_etal_2015,Gagnon_etal_2017}.  Thus the use of the RG, taken as input for SNA  enables us to establish a direct link between noise properties and the predominant chemical pathways traversed by the noise-perturbed network of reactions.
In other words, if one or more of the model parameters ($r$,$\lambda$) run with scale (where the running is controlled by the noise parameters), this
will affect the strengths of the chemical fluxes ($j$) traversing the
pathways, and we then have evidence of noise-controlled fluxes.

\section{Theoretical method}
\label{sec:Theoretical_method}

\subsection{Deterministic CAR model}
\label{sec:CAR_model}

To illustrate our method, we consider a simple well-known spatially
homogeneous cubic autocatalytic reaction (CAR) model (e.g. \cite{Selkov_1968,Gray_Scott_1985}). The
model has an interesting and rich phenomenology when spatially heterogeneous states (rather than only well-stirred, homogeneous states) are considered.  Indeed, simulations of the
deterministic~\cite{Pearson_1993} and
stochastic~\cite{Lesmes_etal_2003} versions of this
reaction-diffusion model reveal the appearance of a variety of
patterns such as stripes, spirals and self-replicating domains. Due
to its autocatalytic nature and the appearance of self-replicating
structures, this model is often taken as an extremely primitive
form of proto-metabolism.

The CAR model involves the following
reactions~\cite{Gray_Scott_1985}:
\begin{eqnarray}
\label{CAR1a}
\mbox{U} + 2\mbox{V} & \stackrel{\lambda}{\rightarrow} & 3\mbox{V}, \\
\label{CAR2a}
\mbox{V} & \stackrel{r_{v}}{\rightarrow} & \mbox{P}, \\
\label{CAR3a}
\mbox{U} & \stackrel{r_{u}}{\rightarrow} & \mbox{Q}, \\
\label{CAR4a}
  & \stackrel{f}{\rightarrow} & \mbox{U}.
\end{eqnarray}
A substrate U, viewed as the ``nutrient'' in the living system
interpretation of this model, is fed into the system at a constant
rate $f$.  The species V, viewed as the ``organism'', consumes the
substrate U and converts it into a copy of V via a second-order
autocatalytic reaction with rate constant~$\lambda$. This
autocatalytic reaction embodies a crude form of proto-metabolism. In
numerical simulations in spatially extended systems, the species V
forms cell-like domains over the substrate U in a certain parameter
range~\cite{Pearson_1993,Mazin_etal_1996,Lesmes_etal_2003,Cooper_etal_2013b}.
Both species V and U decay into inert products P and Q with decay
rates $r_{v}$ and $r_{u}$, respectively.  In a well-stirred system, the
deterministic evolution equations corresponding to reactions~(\ref{CAR1a})-(\ref{CAR4a}) are:
\begin{eqnarray}
\label{eq:Deterministic_CAR_1}
\frac{d V(t)}{d t} & = & - r_{v}V(t) + \lambda U(t)V^{2}(t), \\
\label{eq:Deterministic_CAR_2} \frac{d U(t)}{d t} & = & - r_{u}U(t) -
\lambda U(t)V^{2}(t) + f,
\end{eqnarray}
where $V(t)$ and $U(t)$ represent the time-dependent concentrations of species V and U.

\subsection{Stoichiometric network analysis of the CAR model}
\label{sec:SNA_CAR_model}

\begin{figure}
\includegraphics[width=0.45\textwidth]{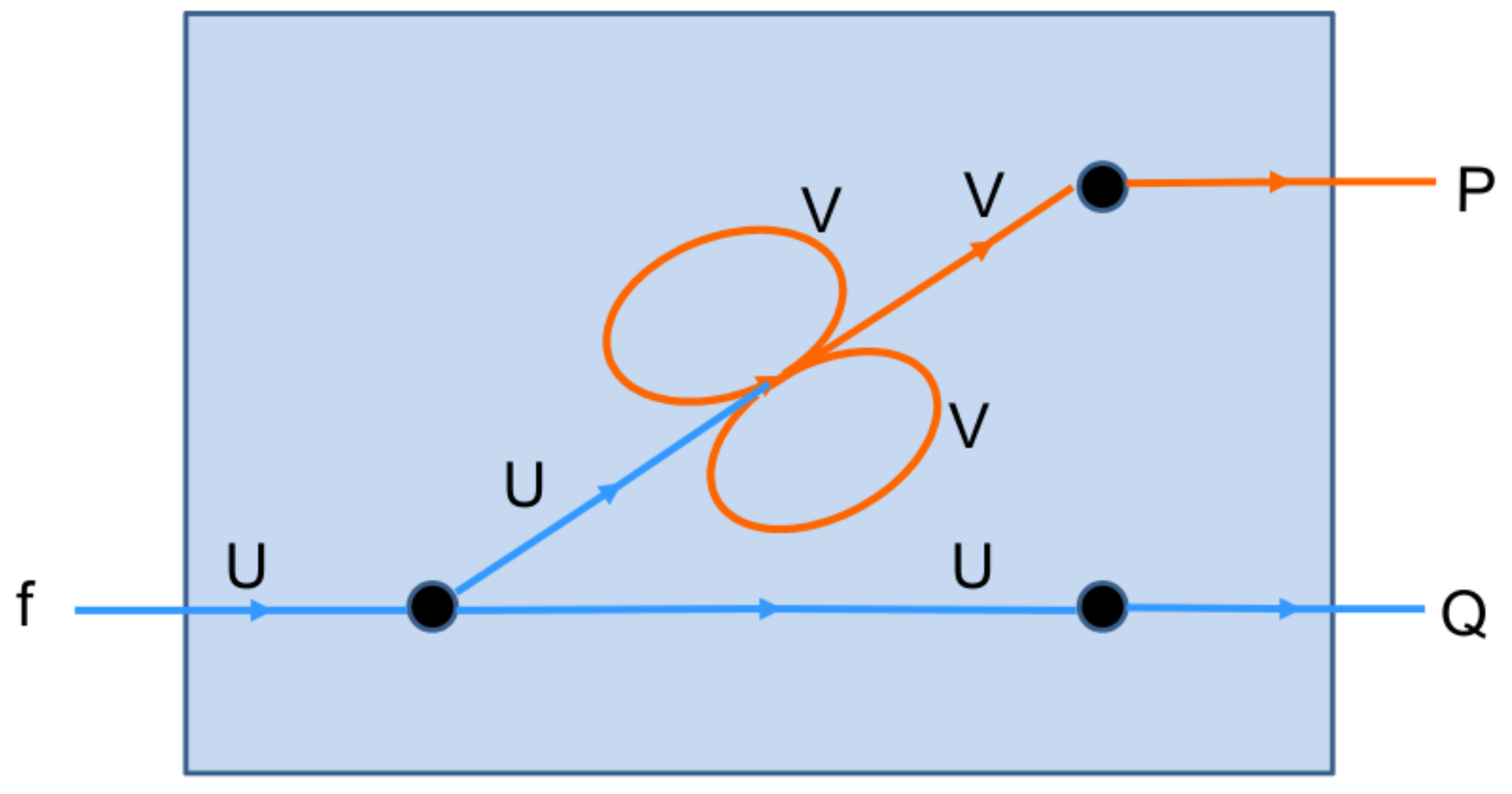}
\caption{The two reaction pathways of the CAR model (see also Table
\ref{TableCAR}). The bounding box encloses a well-mixed system with
input and output flows maintaining the reactions out of equilibrium.
The lower pathway (blue) involves reactions (4) and (3), while
the upper pathway (orange) involves the reactions (4), (1) and
(2); see Table \ref{TableCAR} for details.  An observer
\textit{external} to the enclosure detects only the net
transformations $f\rightarrow Q$ and $f \rightarrow P$.}
\label{fig:Chemical_pathways_CAR}
\end{figure}

The stoichiometric network analysis of the CAR model results as follows (see Appendix~\ref{sec:SNA} for a concise introduction to SNA).  The stoichiometric matrix and extreme flux modes corresponding to the four reactions~(\ref{CAR1a})-(\ref{CAR4a}) are given by :
\begin{eqnarray}
\label{S}
\bm{S} & = & \left[
           \begin{array}{cccc}
             -1& 0 & -1 & 1 \\
             1 & -1 & 0 & 0 \\
           \end{array}
         \right], \\
\label{eq:EFMs}
\bm{E}_1 & = & (0,0,1,1), \;\;\; \bm{E}_2 = (1,1,0,1).
\end{eqnarray}
These EFMs satisfy $\bm{S}\cdot\bm{E}_{1,2} = \bm{0}$, and belong to
the intersection of the right null space of
$\bm{S}$ with the positive orthant $R_{+}^4$.  
These extreme fluxes involve the two elementary chemical pathways of  reactions~(\ref{CAR1a})-(\ref{CAR4a}). These are made explicit
in Table~\ref{TableCAR} and schematically shown in
Fig.~\ref{fig:Chemical_pathways_CAR}.  A general stationary reaction
rate vector $\bm{v}$, for the four reactions, is represented as a
point in $R_{+}^4$, and is expressed as a positive linear
combination of these EFMs:
\begin{equation}\label{va}
\bm{v} = j_1 {\bm E}_1 + j_2 {\bm E}_2 = (j_2,j_2,j_1,j_1+j_2).
\end{equation}
The expansion coefficients $j_i>0$ are the convex parameters, and
correspond to the \textit{magnitudes of the matter fluxes} along the
specific reaction pathway represented by $\bm{E}_i$. As we
demonstrate below, these fluxes can undergo
\textit{renormalization} due to external noise. From
Eqs.~(\ref{CAR1a})-(\ref{CAR4a}) we write the individual stationary
state $(ss)$ reaction rates as a four-component vector:
\begin{equation}\label{vb}
\bm{v} = (\lambda [U]_{ss}[V]_{ss}^2,r_v[V]_{ss},r_u[U]_{ss},f).
\end{equation}
Equating the above two stationary state vectors (\ref{va}) and
(\ref{vb}) and introducing the stationary inverse concentrations
$h_{1}$, $h_{2}$ (where $h_1=h_u = 1/[U]_{ss}$ and $h_2=h_v =
1/[V]_{ss}$) implies:
\begin{equation}\label{CAR1}
\lambda = j_2 h_u h^2_v, \,\,\, r_v = j_2 h_v, \,\,\,
r_u = j_1 h_u, \,\,\, f  = j_1 + j_2.
\end{equation}
Below we use the above identities to deduce the scale-dependent running of
the SNA parameters ($j_{1}$, $j_{2}$, $h_{1}$, $h_{2}$) in terms of
the running of the CAR model parameters ($r_u$, $r_v$, $\lambda$, $f$).

\begin{table}[h]
\begin{tabular*}{\columnwidth}{l @{\extracolsep{\fill}} lllll}
\toprule EFM$\;$ & Reactions & Pathway & Internal species & Net reaction \\
\hline $\bm{E}_1$ & (4) & $f \rightarrow U$ & $U$ & $f\rightarrow Q$  \\
                  & (3)      &  $U \rightarrow Q$ &  \\
                  \hline
$\bm{E}_2$ & (4) & $f \rightarrow U$  & $V,U$ & $f\rightarrow P$ \\
           & (1) & $U + 2V \rightarrow 3V$  & &\\
           & (2)  & $V \rightarrow P$   & & \\
           \bottomrule
\end{tabular*}
\caption{Elementary flux modes (EFM) for the well-mixed CAR model (see
Eqs.~(\ref{CAR1a})-(\ref{CAR4a})) and their corresponding reaction
pathways and internal species. The magnitude of the matter flux
along each pathway is given by the corresponding convex parameter:
$j_1>0, j_2>0$, see Eq.~\ref{va}.} \label{TableCAR}
\end{table}

%

\subsection{Renormalization of the stochastic CAR model}
\label{sec:RG_CAR_model}

To study the effect of external noise on chemical pathways, we add
noise terms $\eta_{v}(t)$, $\eta_{u}(t)$ to the deterministic
equations~(\ref{eq:Deterministic_CAR_1})-(\ref{eq:Deterministic_CAR_2}).
For illustrative purposes, we choose a noise that is widespread in
nature, namely power-law
noise~\cite{PerezMercader_2002,Newman_2005} obeying the following statistics (in Fourier space):
\begin{eqnarray}
\label{eq:Noise_property_1}
\langle \eta_{v}(\omega)\eta_{v}(\omega')\rangle & = & 2A_{v} |\omega/\omega_{v}|^{-\theta_{v}} (2\pi)\delta(\omega+\omega'), \\
\label{eq:Noise_property_2}
\langle \eta_{u}(\omega)\eta_{u}(\omega')\rangle & = & 2A_{u} |\omega/\omega_{u}|^{-\theta_{u}} (2\pi)\delta(\omega+\omega'),
\end{eqnarray}
with all other moments zero.  The amplitudes $A_{v}$, $A_{u}$, exponents $\theta_{v}$, $\theta_{u}$, and inverse time scales $\omega_{v}$, $\omega_{u}$  are free parameters of the noise that can be adjusted experimentally.  Note that other types of experimentally adjustable noise could also be envisaged.

The addition of fluctuations to the CAR model translates into a
non-trivial scaling (or ``running'') of its parameters. The
renormalization group allows us to compute this non-trivial scaling
(see for example
Refs.~\cite{Medina_etal_1989,Tauber_2014} for
the application of RG to stochastic processes).  Note that the
exponents $\theta_{v}$, $\theta_{u}$ are fixed by external
experimental conditions, leading to loop integrals with various
divergence structures. Thus caution must be exercised when applying
dimensional regularization to loop integrals involving power-law
noise terms with arbitrary power-law exponents. Here we follow the
program developed in
Refs.~\cite{Gagnon_PerezMercader_2017,Gagnon_etal_2015,Gagnon_etal_2017}
to compute the running of the stochastic CAR model's parameters with
scale.  

For the purpose of illustration, we focus on the regime where $-3/2
< \theta_{u,v} \leq -1$.  In this regime, and at one-loop order,
only $r_{u}$ and $r_{v}$ develop a logarithmic divergence and run
with scale.  Details of the computation are shown in Appendices~\ref{sec:Feynman_rules}--\ref{sec:RG_calculation}.  The result for $r_{u}(T)$ is (a similar but more complicated expression for $r_{v}(T)$ can be found in Appendix~\ref{sec:RG_calculation}): 
\begin{equation}
\label{eq:ru_regime1}
r_{u}(T) = \left(r_{u}(T^{*}) + \frac{4\lambda A_{v}K_{1}}{|\delta| \omega_{v}^{-\theta_{v}}}\right)\left(\frac{T}{T^{*}}\right)^{|\delta|} - \frac{4\lambda A_{v}K_{1}}{|\delta|\omega_{v}^{-\theta_{v}}}, 
\end{equation}
where $K_{1} = 1/[(4\pi)^{(\theta_{v} + 2)/2}\Gamma((\theta_{v} +
2)/2)]$, $\delta$ is the distance from the logarithmic pole, and
$r_{u}(T^{*})$ is a known value of
the decay rate at some reference temporal scale $T^{*}$.  

The RG analysis allows us to make the following points. The
deterministic CAR
model~(\ref{eq:Deterministic_CAR_1})-(\ref{eq:Deterministic_CAR_2})
exhibits various behaviors (stable solutions, oscillatory solutions,
etc~\cite{Gray_Scott_1983,Gray_Scott_1984,Gray_Scott_1985})
depending on the values of its parameters ($r_{u}$, $r_{v}$,
$\lambda$, $f$).  Adding noise alters the behavior of the
deterministic CAR model, and the renormalization group allows us to
assess quantitatively the extent of this change (provided
perturbation theory is valid, and that no ``new chemistry'' is
encountered  as the temporal scale $T$ is
varied~\footnote{If ``new chemistry'' comes into play at shorter temporal scales, then new species and interactions must be added to Eqs.~(\ref{eq:Deterministic_CAR_1})-(\ref{eq:Deterministic_CAR_2}).  These new interactions might add corrections to the running of the parameters (c.f. Eq.~(\ref{eq:ru_regime1})) that are suppressed by the small temporal scale of the new chemistry~\cite{Gagnon_PerezMercader_2017}.}). In practice, adding noise to the CAR model makes its reaction rate and decay constants
dependent on the noise parameters ($A_{u}$, $A_{v}$, $\theta_{u}$,
$\theta_{v}$, $\omega_{u}$, $\omega_{v}$) and the temporal scale $T$.  In other words, noise
converts the deterministic CAR model into an \textit{effective}
deterministic CAR model, with noise and scale dependent parameters~\footnote{Since the CAR model is perturbatively renormalizable at one-loop~\cite{Gagnon_PerezMercader_2017}, adding noise to Eqs.~(\ref{eq:Deterministic_CAR_1})-(\ref{eq:Deterministic_CAR_2}) does not change the form of the original equations~\cite{Hochberg_etal_1999b}.}.

Note that the RG here is run from a large temporal scale $T^{*}$ to smaller temporal scales $T \leq T^{*}$.  The running of model parameters can be controlled experimentally with the noise in the following way.  We first set the noise parameters ($A_{v}$, $A_{u}$, $\theta_{v}$, $\theta_{u}$) to certain values, and choose a large frequency scale $\omega_{v}^{*} = 2\pi/T^{*}$.  The values of the model parameters ($r_{u}$, $r_{v}$, $\lambda$, $f$) for these values of the noise parameters is the starting point of the running in Fig.~\ref{fig:Plots_running_r_j_h} (a).  By experimentally changing the noise parameter $\omega_{v}^{*}$ to a different value $\omega_{v} > \omega_{v}^{*}$, the number of frequency modes contributing to the second order moment in Eq.~(\ref{eq:Noise_property_1}) changes.  This effectively implements the running in Fig.~\ref{fig:Plots_running_r_j_h} (a), going toward smaller values of $T = 2\pi/\omega_{v}$.

\section{Results and discussion}
\label{sec:Results}

Typically model parameters ($r_{u}$, $r_{v}$, $\lambda$, $f$) are not directly observable.  To make contact with experiments (which is the primary goal of this paper), we apply SNA to this effective deterministic model, in order to see how noise affects observable chemical pathways and the fluxes that traverse them.   To do that, we invert Eq.~(\ref{CAR1}) in order to express the SNA
parameters ($j_{1}$, $j_{2}$, $h_{1}$, $h_{2}$) in terms of the CAR
model parameters ($r_{u}$, $r_{v}$, $\lambda$, $f$):
\begin{eqnarray}
\label{eq:Inverse_correspondence_1} j_{1} & = &  \frac{f -
\sqrt{f^{2} - \frac{4r_{u}r_{v}^{2}}{\lambda}}}{2}, \;\;
j_{2} =   \frac{f + \sqrt{f^{2} - \frac{4r_{u}r_{v}^{2}}{\lambda}}}{2}, \\
\label{eq:Inverse_correspondence_3} h_{1} & = &  \frac{2r_{u}}{f -
\sqrt{f^{2} - \frac{4r_{u}r_{v}^{2}}{\lambda}}}, \;\; h_{2}  =
\frac{2r_{v}}{f + \sqrt{f^{2} - \frac{4r_{u}r_{v}^{2}}{\lambda}}}.
\end{eqnarray}
\noindent The sign choice follows from a stability analysis of the steady
state configurations in the CAR model, which imposes the condition
$j_{1} < j_{2}$ (see Appendix~\ref{sec:Stability_condition} for details).

To obtain the running of the convex parameters ($j_{1}$, $j_{2}$)
and the inverse stationary concentrations ($h_{1}$, $h_{2}$) as a
function of $\omega_{v}$, we substitute
Eqs.~(\ref{eq:ru_regime1}) into
Eqs.~(\ref{eq:Inverse_correspondence_1})-(\ref{eq:Inverse_correspondence_3})
(note that at one-loop order and for $-3/2 < \theta_{v} \leq -1$, the
reaction rate $\lambda$ and the feed rate $f$ do not run, see Appendix~\ref{sec:Power_counting}). Some
representative plots are shown in
Fig.~\ref{fig:Plots_running_r_j_h}.

The decay rates ($r_v,r_u$) for V and U run as a function of $\omega_{v}$, as
shown in Fig.~\ref{fig:Plots_running_r_j_h} (a).  As the
noise frequency scale $\omega_{v}$ is increased, $r_v$ increases whereas $r_u$
decreases with respect to their values $r_u(\omega_{v}^{*})$ and $r_v(\omega_{v}^{*})$
measured at some reference frequency scale. Thus, from
the point of view of an effective deterministic CAR model, the
``nutrient'' species U tends to survive longer but the replicating
species V tends to decay now more rapidly. 
Thus the behavior of the chemical system (dictated by its parameters) depends on the characteristic frequency scale $\omega_{v}$ of the noise.  This scale dependence is absent in the absence of external noise.  

The running in the decay constants ($r_{v}$, $r_{u}$) implies a running in the convex
parameters ($j_1$, $j_2$) as shown in Fig.~\ref{fig:Plots_running_r_j_h} (b). 
We see that $j_{2}$ decreases and $j_{1}$ increases as $\omega_{v}$ increases. Thus
according to Eq.~(\ref{va}),  the matter flux
traversing the catalytic pathway $\bm{E}_2$ diminishes whereas the
flux through the ``unproductive'' pathway $\bm{E}_1$ increases as the
noise frequency scale is increased.   From the point of view of an observer
outside the enclosing box in Fig.~\ref{fig:Chemical_pathways_CAR},
increasing $\omega_{v}$ would result in a decrease of P with respect to Q.


There is a critical scale $\omega_{v}^{c}$ at which the two effective fluxes
equalize $j_1(\omega_{v}^{c}) = j_2(\omega_{v}^c)$ and above which they are undefined.
This feature follows from the
relationships~(\ref{eq:Inverse_correspondence_1})-(\ref{eq:Inverse_correspondence_3})
which develop imaginary parts whenever $f^2 <
\frac{4r_ur_v}{\lambda}$. If the effective decay rates at large
frequency scales grow in magnitude such that they overwhelm the feed
term $f$ at or above that scale, then the latter is unable to
maintain the system in a steady state, and so the SNA approach no
longer applies (SNA is only valid for stationary states). In other
words, the two nontrivial fixed points of
Eqs.~(\ref{eq:Deterministic_CAR_1})-({\ref{eq:Deterministic_CAR_2})
(which are equal to $1/h_{1}$ and $1/h_{2}$) become complex when
$f^2 < \frac{4r_ur_v}{\lambda}$ and thus do not correspond to any
real concentrations for which the system is stationary. 
This signals the onset of a chemical instability, where the system
goes from a bistable regime where the system is either ``alive''
($V\neq 0$) or ``dead'' ($V= 0$) to a single stable regime where the
system consists solely of a uniform distribution of nutrient~U.


\begin{figure}[H]
(a)

\parbox{0.5\textwidth}{\includegraphics[width=0.43\textwidth]{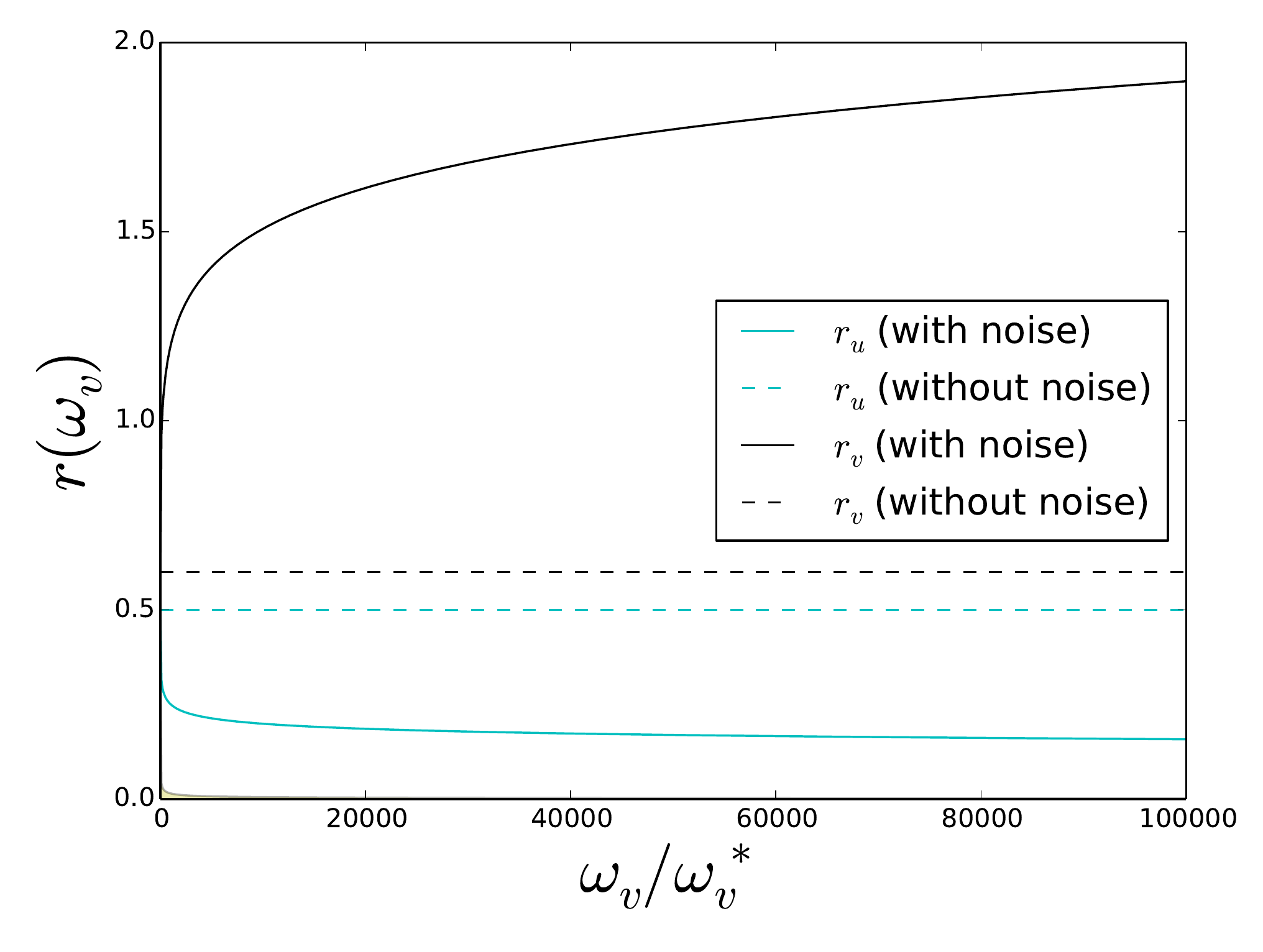}}

(b) 

\parbox{0.5\textwidth}{\includegraphics[width=0.43\textwidth]{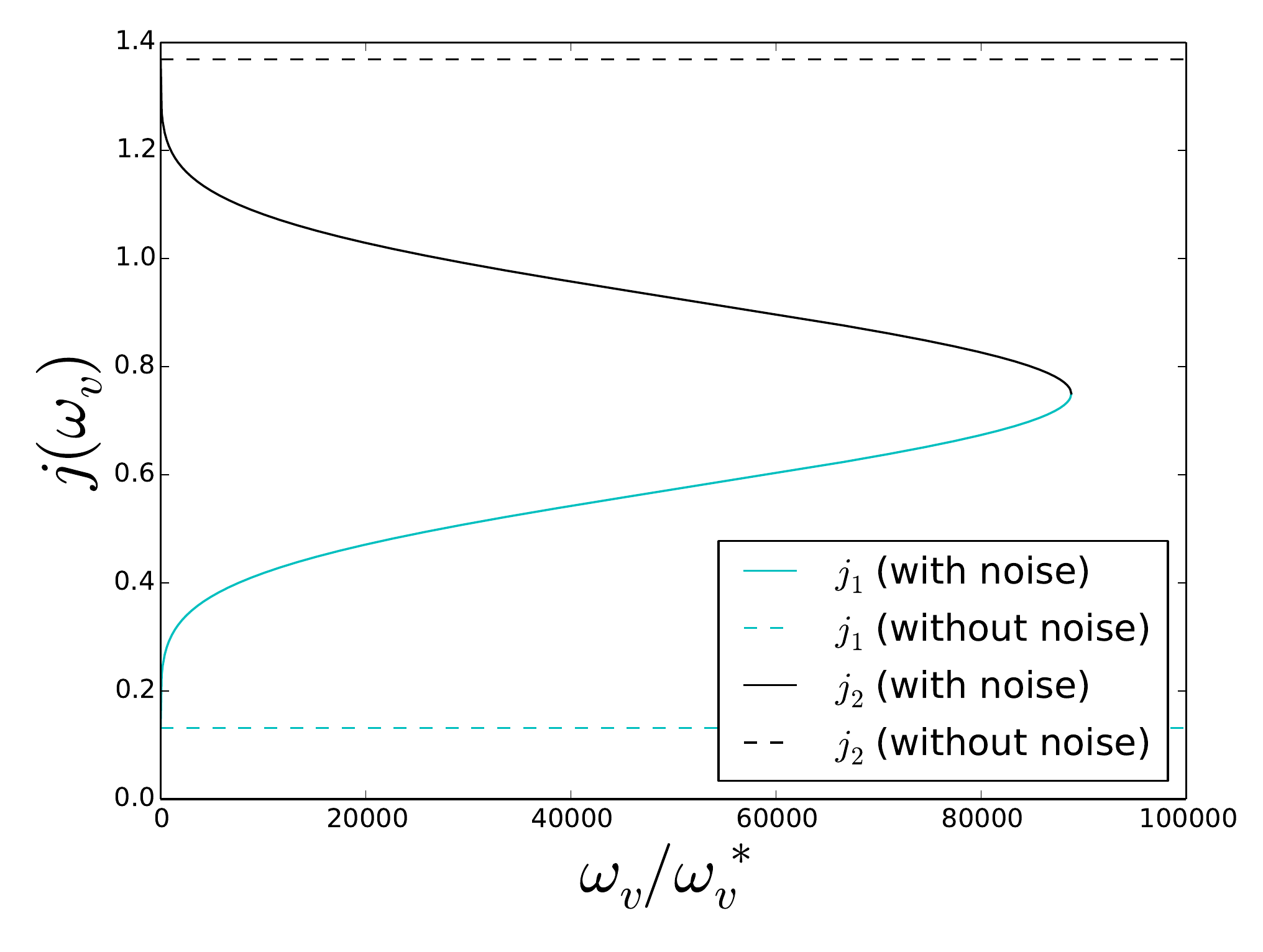}}

(c)

\parbox{0.5\textwidth}{\includegraphics[width=0.43\textwidth]{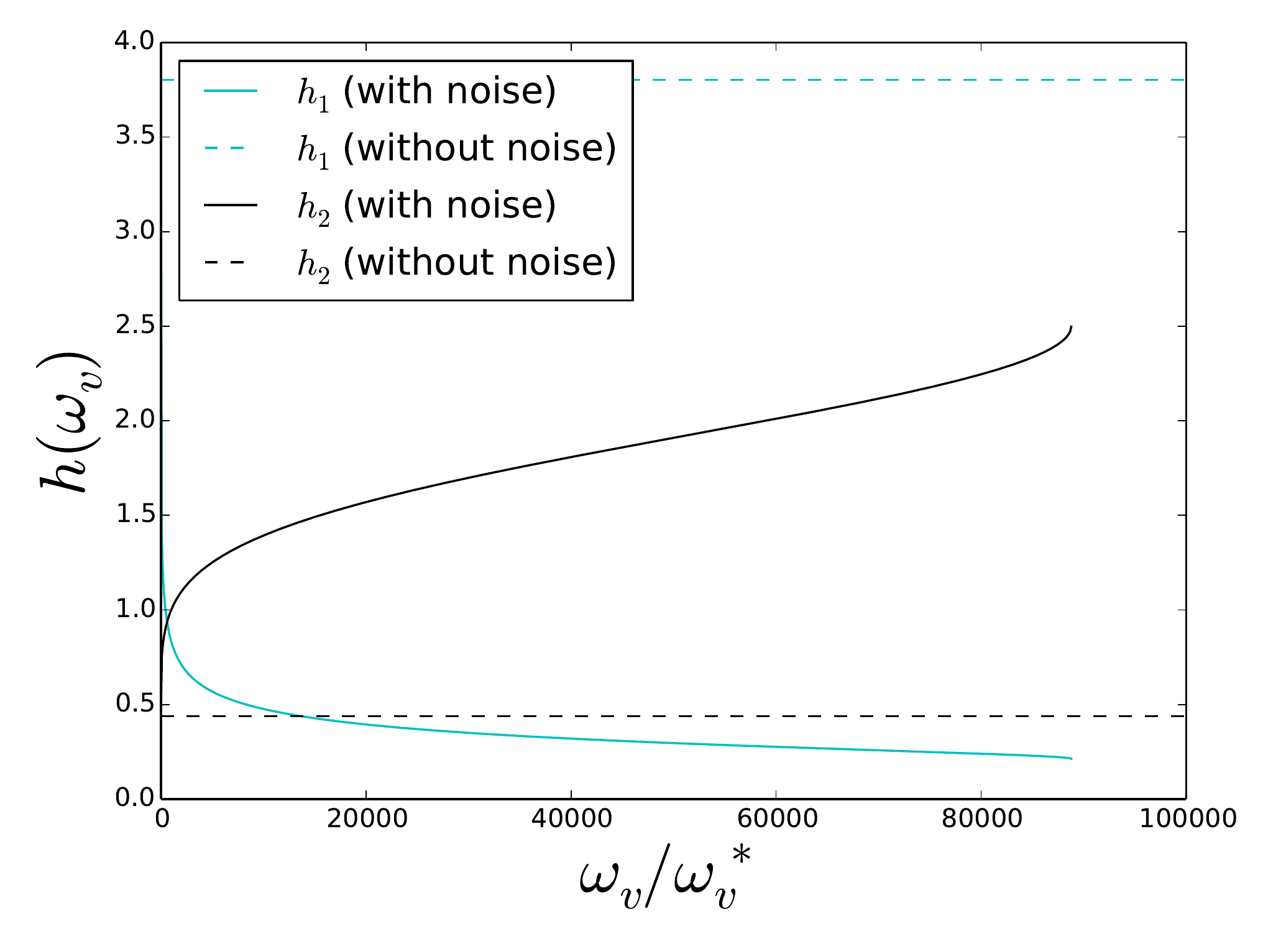}}
\caption{Running of the parameters $r_{u}$, $r_{v}$ (a), $j_{1}$,
$j_{2}$ (b), and $h_{1}, h_{2}$ (c) as a function of
scale.  Fixed parameters are $\omega_{v}^{*} = 0.1$, $r_{u}(T^{*}) = 0.5$,
$r_{v}(T^{*}) = 0.6$, $\lambda = 1.0$, $f=1.5$, $\delta = 0.1$.
Dotted lines represent the running without noise ($A_{v} = 0$,
trivial scaling), while full lines represent the running with noise
($A_{v} = 0.001 \neq 0$).} \label{fig:Plots_running_r_j_h}
\end{figure}

The inverse stationary state concentrations ($h_1,h_2$) also run with the
noise frequency scale $\omega_{v}$ as shown in 
Fig.~\ref{fig:Plots_running_r_j_h} (c). Note here that as $\omega_{v}$ is
increased, the observed decrease in $h_1=1/[V]_{ss}$ corresponds to
an increase in the steady-state concentration of species V. So there
is relatively more V (replicating species) at shorter time scales
(greater concentrations) with respect to the concentration at the
reference frequency scale $\omega_{v}^{*}$. At the scale where the two inverse
concentrations become imaginary, the system goes from a bistable to
a monostable regime and effectively ``dies''. 
In this case, from the point of view of an observer outside the enclosing box in
Fig~\ref{fig:Chemical_pathways_CAR}, decreasing the noise frequency scale
$\omega_{v}$ would result in an increase of P with respect to Q.

\section{Conclusions}
\label{sec:Conclusion}

We have demonstrated that external adjustable noise can be used to
control the directly observable matter fluxes that traverse the reaction pathways in an
overall reaction model by combining the renormalization group with
stoichiometric network analysis. For the case of the CAR model
treated here, the fluxes along the driven autocatalytic pathway
$\bm{E}_2$ and along the driven unproductive flow-through pathway
$\bm{E}_1$ can be controlled. SNA predicts generally that the
renormalization of reaction model parameters implies an associated
renormalization of the convex parameters (the flux magnitudes) and
the inverse stationary concentrations. The feasibility of noise
controlled fluxes is thus expected in general complex reaction
networks coupled to external noise sources, and has recently been reported for the BZ reaction~\cite{Srivastava_etal_2018}.

The physico-chemical interpretation of the results of this paper open the door to the extension and application in many directions where optimization or selection of ``chemical'' pathways is naturally occurring or desirable (as an example, noise control could be used to implement chemical logic gates using the approach in Ref.~\cite{Egbert_etal_2018}).  This is due to the fact that, depending on noise statistics, one can channel energy at the molecular levels to processes where the external energy selectively provided by the noise  makes the system visit some pathways more frequently than others, as opposed to the situation without noise.  A future direction for this research would be to try this technique in more complicated chemical models, where it might be possible to shut down a pathway or activate a previously non-accessible one. This is a direct consequence of the connection shown here between noise parameters and stoichiometry.  Potential practical applications range from electrochemistry, systems chemistry, epidemiology, immunology and ecology to large scale industrial processes and environmental applications.

\section*{Acknowledgements}
J.-S. G. and J. P.-M. thank Repsol S. A. for its support and D. H. acknowledges the project CTQ2017-87864-C2-2-P (MINECO) Spain.

\appendix

\section{Stoichiometric network analysis}
\label{sec:SNA}

We summarize the basic notions of the stoichiometric network
analysis needed in the present paper. A fuller detailed account of
SNA and a concise review are given in Refs.~\cite{Clarke80,Clarke88}. The
chemical reactions for $r$ reactions and $n$ reacting species obeying mass-action kinetics can be
written as:
\begin{equation}\label{reaction}
\alpha_{1j}S_1 + \ldots + \alpha_{nj} S_n \stackrel{k_j}\rightarrow
\beta_{1j}S_1 + \ldots + \beta_{nj} S_n, \;\;\; j=1,\ldots , r,
\end{equation}
where the $S_i$, $1 \leq i \leq n$, are the chemical species and
each $k_j$ the reaction rate constant for the $j$th reaction. From
the coefficients in Eq.~(\ref{reaction}) we construct the $n \times
r$ stoichiometric matrix $\bm{S}$ with elements:
\begin{equation}
S_{ij} = \beta_{ij} - \alpha_{ij}.
\end{equation}
The reaction rate of the $j$-th reaction, assuming mass action
kinetics, takes the form of a monomial:
\begin{equation}\label{monomials}
v_j(x,k_j) = k_j \prod_{i=1}^n x_i^{\kappa_{ij}},
\end{equation}
where $\kappa_{ij}=\alpha_{ij}$ is the molecularity of the species
$S_i$ in the $j$th reaction, $\bm{\kappa}$ the $n \times r$ kinetic
matrix. The $x_i = [S_i]$ denote concentrations and $v_j$ is the
flux or reaction rate of the $j$th reaction.

Dynamic mass balance equations for the system shown in Eq.~(\ref{reaction})
can be written as (in vector notation):
\begin{equation}\label{rateequ}
\frac{d \bm{x}}{dt} = \bm{S} \bm{v}.
\end{equation}
Just as for the stoichiometry, the pathway structure should be an
invariant property of the reaction network. We can find this from
the steady state condition:
\begin{equation}\label{nullspace}
\bm{0} = \bm{S} \bm{v},
\end{equation}
which defines the right null space of $\bm{S}$, and corresponds to
the set of all stationary-state $(ss)$ solutions $(\bm{v})$ of
Eq.~(\ref{rateequ}). Since the reaction rates in
Eq.~(\ref{monomials}) are positive-definite, they satisfy $v_i(x,k)
> 0$, and therefore must belong to the \textit{intersection} of the null
space Eq. (\ref{nullspace}) with the positive orthant $R_{+}^r$:
\begin{equation}
\label{intersection}
v(x_{ss},k) \in \{ z \in {R}^r | Sz = 0, z \in R_{+}^r \} = \ker(S)\bigcap R^r_{+}.
\end{equation}

This intersection defines a convex polyhedral cone $C_v$
\cite{Rockafeller} spanned by a set of $M$ minimal generating
vectors $\bm{E}_i$'s (see Fig.~\ref{fig:Convex Cone(a)}):
\begin{equation}\label{convex}
C_v = \{ \bm{v}=\sum_{i=1}^M j_i \bm{E}_i : j_i > 0 \}.
\end{equation}
These extreme currents or extreme flux modes (EFM) $\{ \bm{E}_i
\}_{i=1}^M$ are vectors having $r$-components, equal to the number
of reactions~\cite{Clarke88}.  The positive definite expansion
coefficients $j_i > 0$ are called the convex parameters. Programs,
such as COPASI, are freely available for calculating these extreme
currents~\cite{Copasi}.

Define $h_i =1/(x_{ss})_i$ as the inverse of the stationary state
concentration. Then conventional reaction rate constants can be
written in terms of the SNA variables $(j,h)$ through the
identities~\cite{Clarke88}:
\begin{equation}\label{krates}
k_l = \Big(\sum_{i=1}^M j_i \bm{E}_i \Big)_l
\prod_{i=1}^n(h_i)^{\kappa_{il}} \;\; \Rightarrow \;\;
k_l=k_l(j,h).
\end{equation}
This follows immediately from Eqs.~(\ref{monomials}) and (\ref{convex}) and the
definition of $h_i$. This general relation relates running in the
reaction rate constants $(k)$ to running in the SNA variables
$(j,h)$.

\begin{figure}[h]
\includegraphics[width=0.40\textwidth]{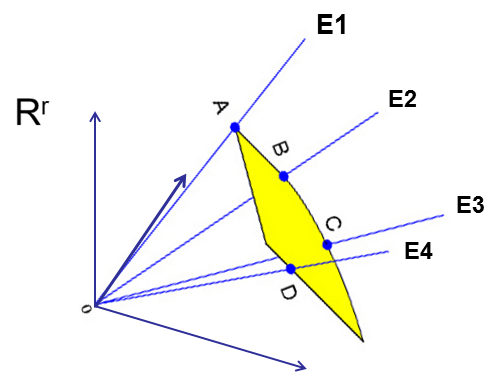}
\caption{The convex cone $C_v$, where each orthogonal axis in the positive
orthant $R^r_{+}$ corresponds to one of the $1 \leq j \leq r$
\textit{stationary} reaction rates $v_j$ (see Eq.~(\ref{monomials})) and so
satisfies Eq.~(\ref{nullspace}) and belongs to the intersection Eq.~(\ref{intersection}). The general stationary reaction rate can be
written in vector form, as a point in this cone: a linear
combination of the $M$ cone edge vectors $\bm{E}_i$ with positive
coefficients $j_i >0$. For purposes of clarity, only four such edge
vectors are drawn here.} \label{fig:Convex Cone(a)}
\end{figure}
%

\section{Feynman rules for the CAR model}
\label{sec:Feynman_rules}

Following standard procedures~\cite{Medina_etal_1989,Hochberg_etal_2003,Gagnon_PerezMercader_2017}, the Feynman rules for the CAR model can be obtained from the formal solution of the CAR evolution equations.  The various building blocks are shown graphically in Fig.~\ref{fig:Feynman_rules_CAR}, where the free response functions are given by:
\begin{eqnarray}
\label{eq:Free_response_function_U}
G_{u0}(\omega) & = & \frac{1}{- i\omega + r_{u}}, \\
\label{eq:Free_response_function_V}
G_{v0}(\omega) & = & \frac{1}{- i\omega + r_{v}},
\end{eqnarray}
the noise insertions by:
\begin{eqnarray}
N_{u0}(\omega) & = & 2A_{u}\left|\omega/\omega_{u}\right|^{-\theta_{u}}, \\
N_{v0}(\omega) & = & 2A_{v}\left|\omega/\omega_{v}\right|^{-\theta_{v}},
\end{eqnarray}
and the vertices by:
\begin{eqnarray}
\label{eq:4_vertex}
\Gamma_{u0} & = & -\Gamma_{v0} \;=\; -\lambda, \\
\label{eq:3_vertex}
\Gamma_{u0}^{(f)} & = & -\Gamma_{v0}^{(f)} \;=\; -\frac{\lambda f}{r_{u}}.
\end{eqnarray}
To write down a specific Feynman diagram, the above graphical rules must be supplemented with conservation of frequency at each vertex and integration over undetermined frequencies.  Note that to obtain the vertices in Eq.~\ref{eq:3_vertex}, we use the re-definition $U \rightarrow \tilde{U} = U - f/r_{u}$ to get rid of the constant feeding term.
\begin{figure}[h]
\includegraphics[width=0.47\textwidth]{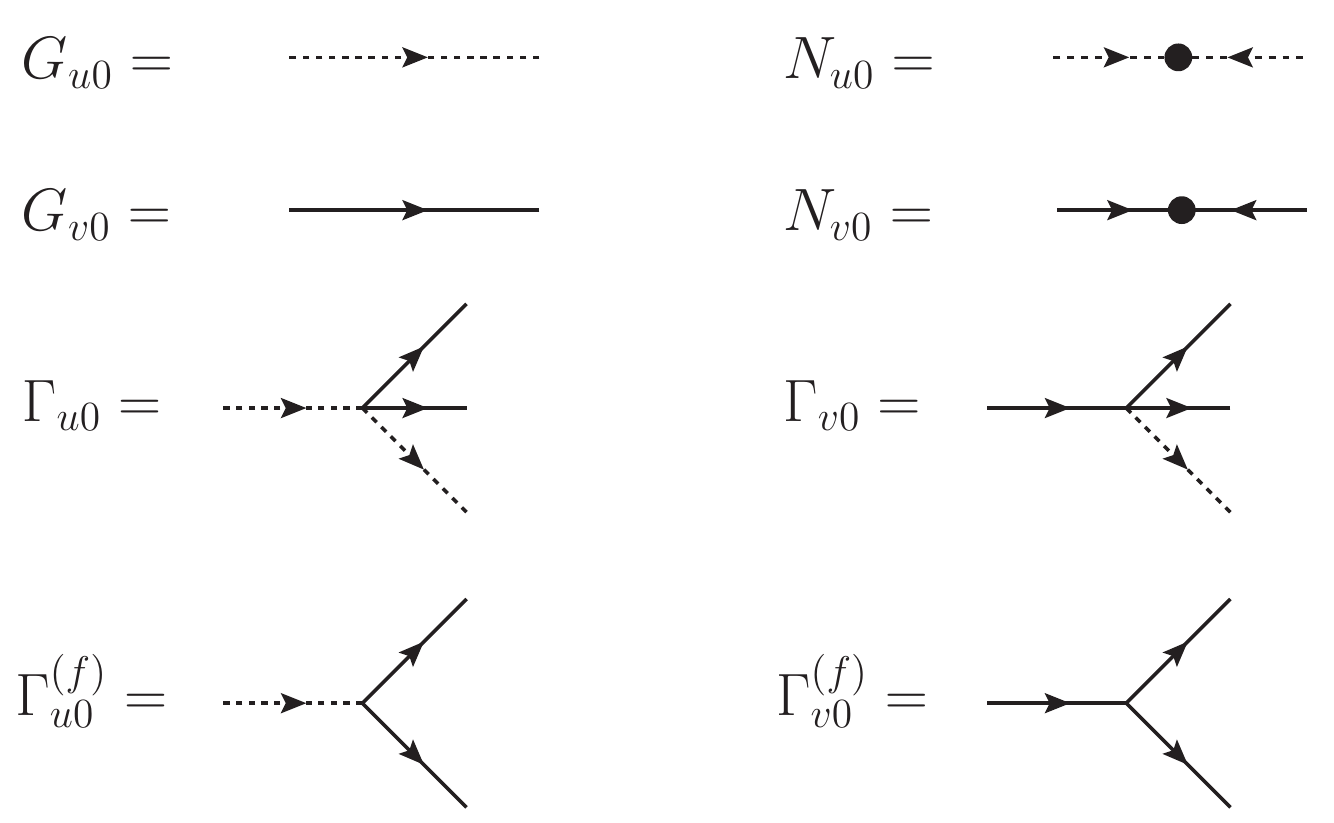}
\caption{Feynman rules for the CAR model.  See text for definitions of symbols.}
\label{fig:Feynman_rules_CAR}
\end{figure}
%

\section{Power counting}
\label{sec:Power_counting}

In this paper, we restrict ourselves to one-loop calculations, and we are interested in diagrams that diverge in the ultraviolet (UV).  For power counting purposes, response functions and noise insertions can be estimated as follows in the UV limit:
\begin{eqnarray}
G_{v0} \;\sim\; G_{u0} & \sim & \omega^{-1}, \\
N_{v0} \;\sim\; N_{u0} & \sim & A_{u,v} \omega^{-\theta_{u,v}},
\end{eqnarray}
and each loop integration contributes one power of $\omega$.  Using the above, we can find the UV divergence structure of corrections to various CAR model parameters.  Figure~\ref{fig:Response_functions_one_loop} shows one-loop diagrams corresponding to corrections to $G_{u0}$ and $G_{v0}$ (or $r_{u}$ and $r_{v}$).  Power counting gives $\Gamma_{r_{u}}^{\rm (row\;1)} \sim \Gamma_{r_{v}}^{\rm (row\;2,right)} \sim \Lambda^{-\theta_{v}-1}$ and $\Gamma_{r_{v}}^{\rm (row\;2,left)} \sim \Lambda^{-\theta_{v}-2}$, where $\Lambda$ is a large frequency cutoff scale.       One-loop corrections to $\Gamma_{u0}$ and $\Gamma_{v0}$ (or $\lambda$) are shown in Fig.~\ref{fig:4_point_vertex}.  Counting powers of frequency for each diagram, we get that $\Gamma_{u}^{\rm (row\; 1)} \sim -\Gamma_{v}^{\rm (row\; 1)} \sim \Lambda^{-\theta_{v}-2}$ and $\Gamma_{u}^{\rm (row\; 2)} \sim -\Gamma_{v}^{\rm (row\; 2)} \sim \Gamma_{u}^{\rm (row\; 3)} \sim -\Gamma_{v}^{\rm (row\; 3)} \sim \Lambda^{-\theta_{v}-3}$.         One-loop corrections to $\Gamma_{u0}^{(f)}$ and $\Gamma_{v0}^{(f)}$ (or $f$) are shown in Fig.~\ref{fig:3_point_vertex}.  Power counting gives $\Gamma_{u}^{(f)\rm \; (row\;1)} \sim -\Gamma_{v}^{(f)\;\rm (row\;1)} \sim \Lambda^{-\theta_{v} - 2}$ and $\Gamma_{u}^{(f)\rm \; (row\;2)} \sim -\Gamma_{v}^{(f)\;\rm (row\;2)} \sim \Lambda^{-\theta_{v} - 3}$.         One-loop corrections to $N_{u0}$ and $N_{v0}$ (or $A_{u}$ and $A_{v}$) are shown in Fig.~\ref{fig:2_point_noise}.  Counting powers of frequency for each diagram, we get that $N_{u} \sim N_{v} \sim \Lambda^{-2\theta_{v} -3}$.

Note that none of the one-loop diagrams shown in Figs.~\ref{fig:Response_functions_one_loop}--\ref{fig:2_point_noise} depend on the noise $\eta_{u}(t)$.  Noise on the U chemical starts to contribute to the running of parameters at two-loop, and is thus negligible compared to the effect of noise on the V chemical.

Each diagrams in Figs.~\ref{fig:Response_functions_one_loop}--\ref{fig:2_point_noise} may be UV divergent, depending on the noise exponent $\theta_{v}$.  From the above power counting, we can identify three regimes:

\begin{description}
	\item[Regime 1]  When $-\frac{3}{2} < \theta_{v} \leq -1$, two parameters ($r_{u}$ and $r_{v}$) run.

	\item[Regime 2] When the temporal noise exponent is $-2 < \theta_{v} \leq -\frac{3}{2}$, four parameters ($r_{u}$, $r_{v}$, $A_{u}$, $A_{v}$) run.  There is even a chance that the exponents themselves ($\theta_{u}$, $\theta_{v}$) might also run, depending on the form of the one-loop corrections.

	\item[Regime 3]  When the temporal noise exponent is $\theta_{v} \leq -2$, six parameters ($r_{u}$, $r_{v}$, $A_{u}$, $A_{v}$, $\lambda$, $f$) run.  For low enough values of $\theta_{v}$, non-renormalizable operators might also play an important role in the dynamics, as discussed in Ref.~\cite{Gagnon_PerezMercader_2017}.
\end{description}

For simplicity and for the purpose of illustration, we concentrate on Regime 1 in the following.

\begin{figure}[h]
\includegraphics[width=0.46\textwidth]{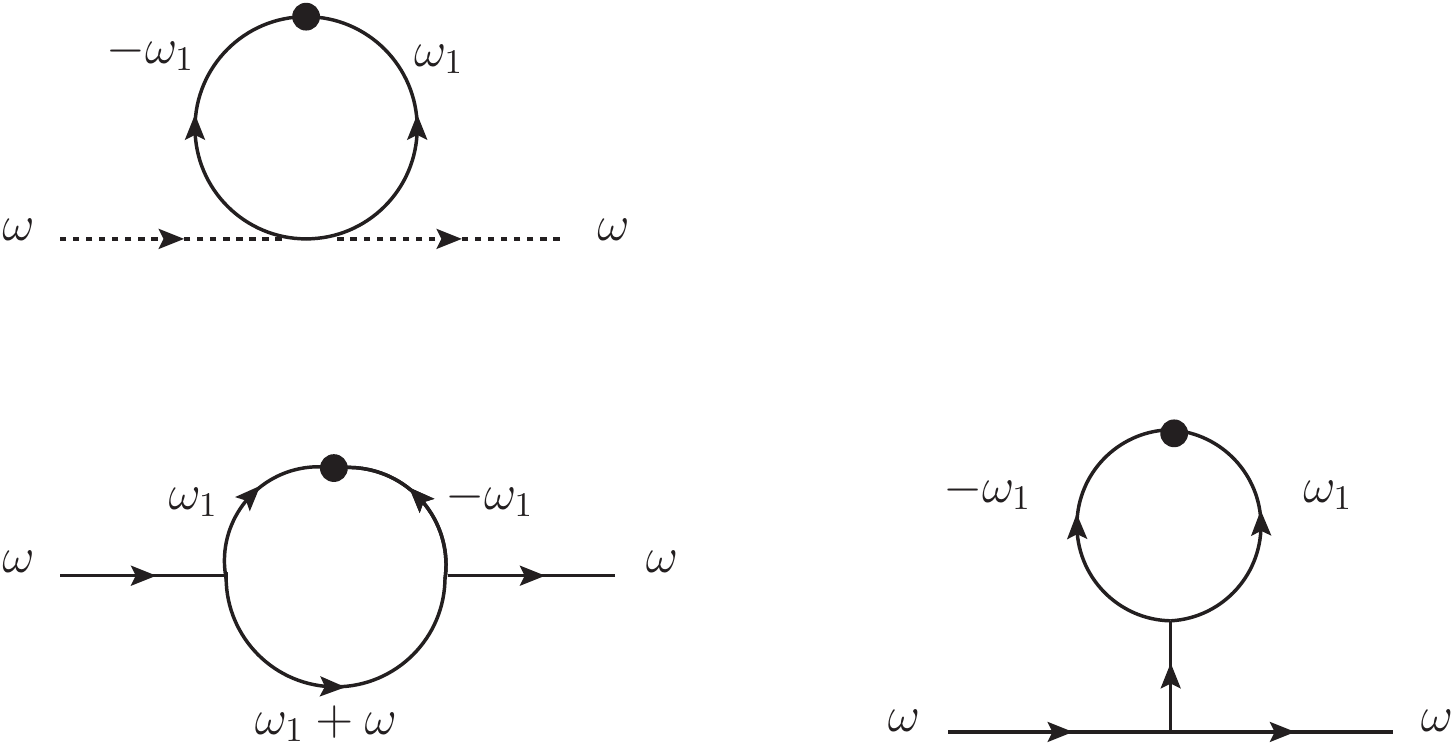}
\caption{One-loop corrections to $G_{u0}$ (top row) and $G_{v0}$ (lower row).}
\label{fig:Response_functions_one_loop}
\end{figure}
\begin{figure}[h]
\includegraphics[width=0.40\textwidth]{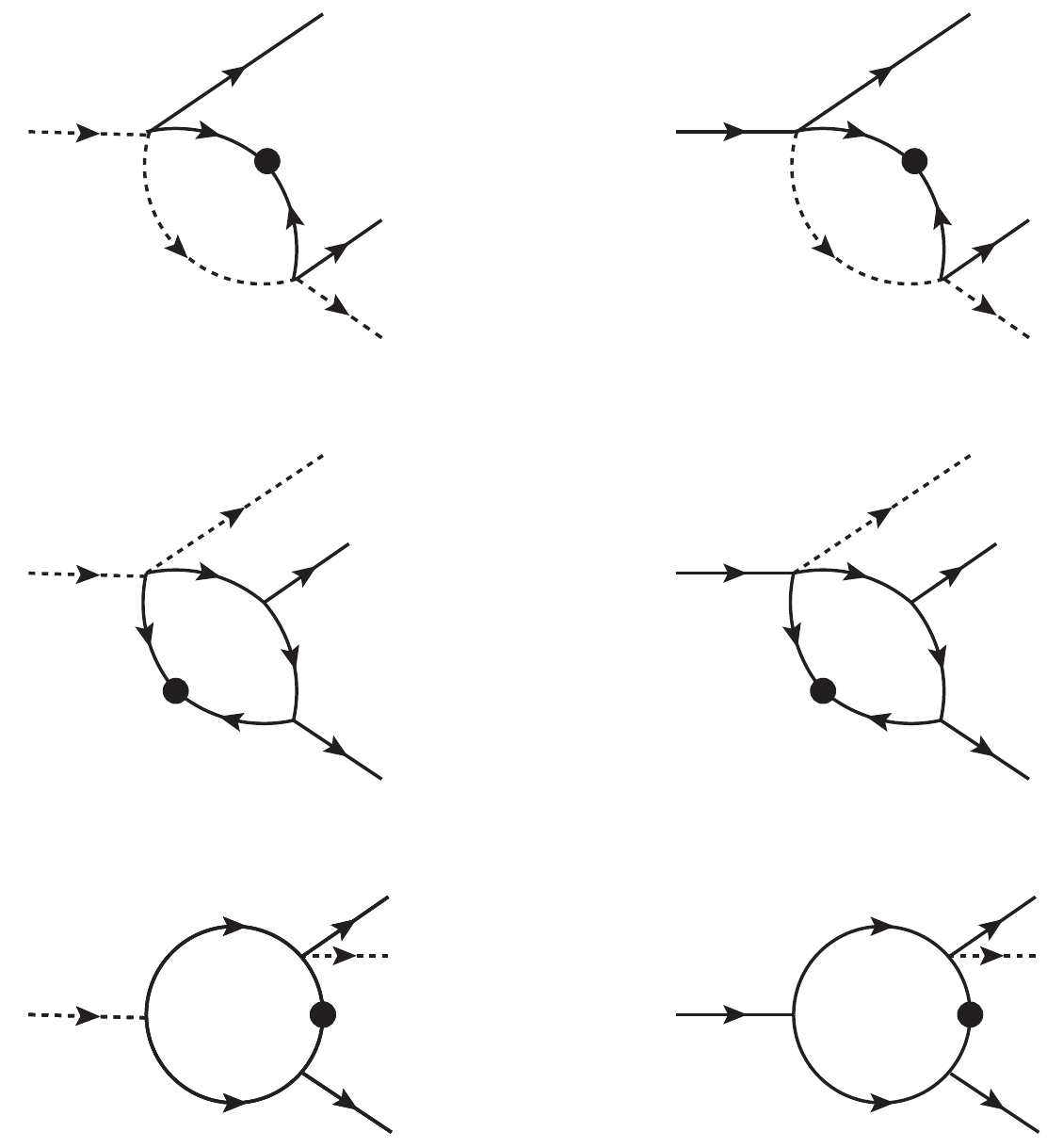}
\caption{One-loop corrections to $\Gamma_{u0}$ (left column) and $\Gamma_{v0}$ (right column).}
\label{fig:4_point_vertex}
\end{figure}
\begin{figure}[h]
\includegraphics[width=0.40\textwidth]{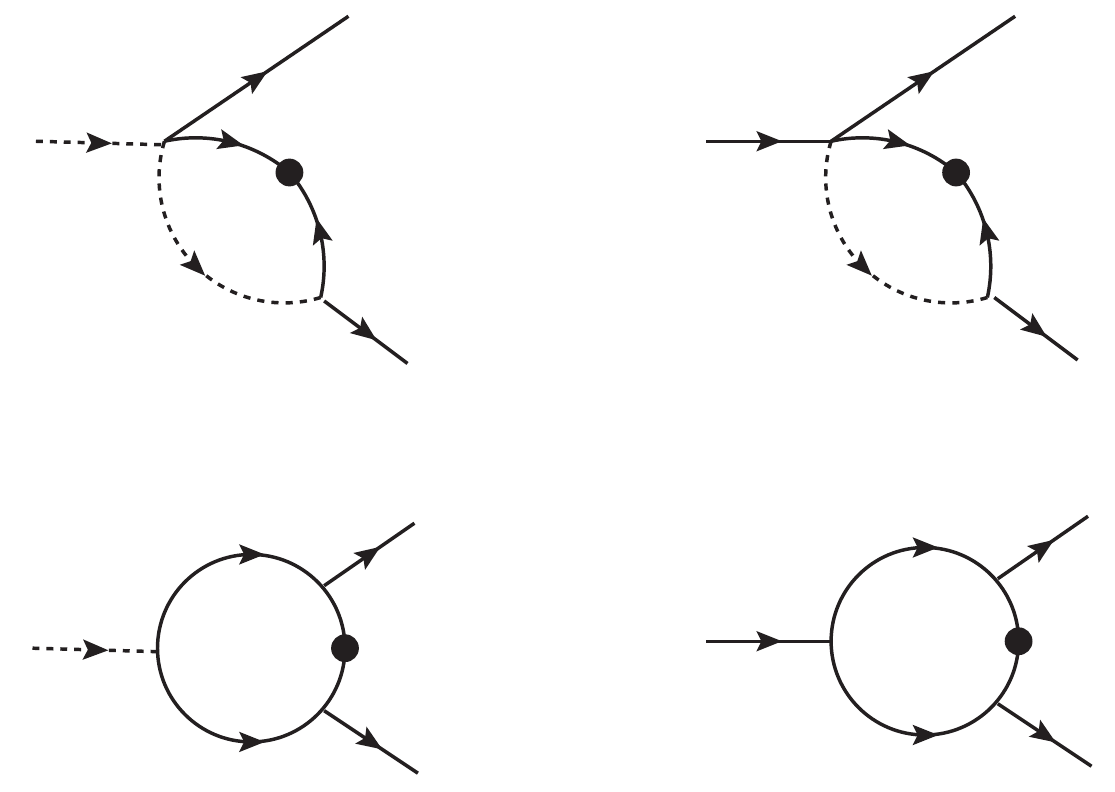}
\caption{One-loop corrections to $\Gamma_{u0}^{(f)}$ (left column) and $\Gamma_{v0}^{(f)}$ (right column).}
\label{fig:3_point_vertex}
\end{figure}
\begin{figure}[h]
\includegraphics[width=0.45\textwidth]{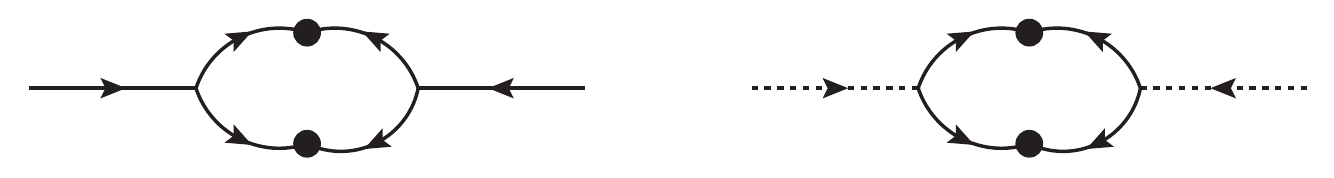}
\caption{One-loop corrections to $N_{v0}$ (left column) and $N_{u0}$ (right column).}
\label{fig:2_point_noise}
\end{figure}
%

\section{Renormalization group flow of the model parameters}
\label{sec:RG_calculation}

In this section, we present some details on how to obtain the running of the model parameter $r_{u}$ ($r_{v}$ is done in a similar way) by computing its $\beta$-function.  We start with the one-loop correction to the response function $G_{u0}$ (see Fig.~\ref{fig:Response_functions_one_loop}, top row):
\begin{eqnarray}
\Gamma_{r_{u}}(\omega') & = & - \frac{2\lambda A_{v}}{\omega_{v}^{-\theta_{v}}} \int\frac{d\omega}{(2\pi)}\;  |\omega|^{-\theta_{v}} \; G_{v0}(\omega)G_{v0}(-\omega), \nonumber \\
                        & = & - \frac{2\lambda A_{v}}{\omega_{v}^{-\theta_{v}}} \int\frac{d\omega}{(2\pi)}\;  |\omega|^{-\theta_{v}} \; \left(\frac{1}{\omega^{2} + r_{v}^{2}}\right).
\end{eqnarray}
To regulate this potentially divergent integral, we analytically continue the time dimension to $z$:
\begin{equation}
\label{eq:One_loop_integral_ru_2}
\Gamma_{r_{u}}(\omega') = - \frac{2\lambda A_{v}^{(z)}}{\omega_{v}^{-\theta_{v}}} \int\frac{d^{z}\omega}{(2\pi)^{z}}\;  |\omega|^{-\theta_{v}} \; \left(\frac{1}{\omega^{2} + r_{v}^{2}}\right).
\end{equation}
where the superscript $(z)$ in $A_{v}^{(z)}$ indicates that the engineering dimension of the noise amplitude depends on the analytically continued time dimension $z$.  The integral in Eq.~(\ref{eq:One_loop_integral_ru_2}) can be done using the method regularization in the presence of noise of Ref.~\cite{Gagnon_etal_2015}.  The result is:
\begin{equation}
\label{eq:One_loop_integral_ru_3}
\Gamma_{r_{u}}(\omega') = - \frac{2\lambda A_{v}^{(z)}}{\omega_{v}^{-\theta_{v}}} \frac{\pi}{(4\pi)^{z/2}\Gamma(z/2)} \; \frac{(r_{v})^{-2+z-\theta_{v}}}{\sin \pi\left(\frac{z}{2}-\frac{\theta_{v}}{2}\right)}.
\end{equation}
Equation~(\ref{eq:One_loop_integral_ru_3}) has an infinite number of poles, and the location of those poles depend on the noise exponent $\theta_{v}$.  Focusing on Regime 1, we expand Eq.~(\ref{eq:One_loop_integral_ru_3}) around the pole located at $\theta_{v} = -1$, corresponding to a logarithmic UV divergence.  Defining the quantity $z - \theta_{v} = 2 - \delta$ for convenience and performing a $\delta$-expansion of Eq.~(\ref{eq:One_loop_integral_ru_3}), we obtain:
\begin{eqnarray}
\label{eq:One_loop_integral_ru_4}
\Gamma_{r_{u}}(\omega') & = & -\frac{4\lambda A_{v}^{(\theta_{v}+2)}K_{1}}{\omega_{v}^{-\theta_{v}}} \frac{1}{\delta} + \mbox{finite},
\end{eqnarray}
where $K_{1} = 1/[(4\pi)^{(\theta_{v} + 2)/2}\Gamma((\theta_{v} +
2)/2)]$ and ``finite'' means terms that are finite in the $\delta\rightarrow 0$ limit.  Those terms are not necessary for $\beta$-function computations, and we ignore them in the following.

The Z-factor for $r_{u}$ is given by:
\begin{eqnarray}
\label{eq:Z_factor_ru}
Z_{r_{u}} & = & 1 + \frac{\Gamma_{r_{u}}(\omega')}{r_{u}} \;=\; 1 - \frac{4g_{u}^{(z)}K_{1}T^{\delta}}{\delta},
\end{eqnarray}
where $T$ is an arbitrary temporal scale and where we defined the effective coupling:
\begin{eqnarray}
\label{eq:Effective_coupling_gu}
g_{u}^{(z)} & = & \frac{\lambda A_{v}^{(z)}}{r_{u}}.
\end{eqnarray}
The $\beta$-function for $r_{u}$ is obtained by taking the derivative of the bare effective coupling~(\ref{eq:Effective_coupling_gu}) with respect to the arbitrary timescale $T$.  The result is:
\begin{eqnarray}
\label{eq:Beta_function_gu}
\beta_{g_{u}} & \equiv & T\frac{dg_{u}}{dT} \;=\;  \delta g_{u} - 4g_{u}^{2}K_{1} + O(g^{3}).
\end{eqnarray}
In terms of the original model parameters, the $\beta$-function becomes:
\begin{eqnarray}
\beta_{r_{u}} & \equiv & T\frac{dr_{u}}{dT} \;=\; -\delta r_{u} + \frac{4\lambda A_{v} K_{1}}{\omega_{v}^{-\theta_{v}}}.
\end{eqnarray}
Integrating the above $\beta$-function gives the running of $r_{u}$ as a function of the arbitrary temporal scale $T$:
\begin{equation}
\label{eq:ru_regime1_appendix}
r_{u}(T) = \left(r_{u}(T^{*}) + \frac{4\lambda A_{v}K_{1}}{|\delta| \omega_{v}^{-\theta_{v}}}\right)\left(\frac{T}{T^{*}}\right)^{|\delta|} - \frac{4\lambda A_{v}K_{1}}{|\delta|\omega_{v}^{-\theta_{v}}}, 
\end{equation}
where $r_{u}(T^{*})$ is an experimentally known value of
the decay rates at some reference temporal scale $T^{*}$.  The expression for $r_{v}(T)$ can be obtained in a similar way:
\begin{eqnarray}
\label{eq:rv_regime1_appendix}
r_{v}(T) & = & \left(\frac{T}{T^{*}}\right)^{-|\delta|} \left[r_{v}^{2}(T^{*}) + \frac{\frac{16\lambda^{2}f^{2}A_{v}K_{1}}{|\delta| \omega_{0}^{-\theta_{v}}} \ln\left(\frac{r_{u}(T^{*})}{r_{u}(T)}\right)}{\left(r_{u}(T^{*}) + \frac{4\lambda A_{v}K_{1}}{|\delta| \omega_{0}^{-\theta_{v}}}\right)^{2}} \right. \nonumber \\
         &   & \left. - \frac{\frac{64\lambda^{3}f^{2}A_{v}^{2}K_{1}^{2}}{|\delta|^{2} \omega_{0}^{-2\theta_{v}}}\left(\frac{1}{r_{u}(T^{*})} - \frac{1}{r_{u}(T)}\right)}{\left(r_{u}(T^{*}) + \frac{4\lambda A_{v}K_{1}}{|\delta| \omega_{0}^{-\theta_{v}}}\right)^{2}} \right]^{1/2}.
\end{eqnarray}

\section{Stability analysis of steady states}
\label{sec:Stability_condition}

Stability analysis is carried out in terms of the positive convex
parameters $j_i > 0$ and the positive inverse stationary concentrations
$h_i > 0$. The Jacobian matrix is given by~\cite{Gatermann}:
\begin{eqnarray}\label{Jacobian}
{\rm Jac}(\bm{j},\bm{h}) &=& \bm{S} {\rm diag}(\bm{j \cdot E})
\bm{\kappa^T}{\rm diag}(\bm{h}).
\end{eqnarray}
Substituting in Eqs (\ref{S})-(\ref{eq:EFMs}) this gives:
\begin{equation}
{\rm Jac}(\bm{j},\bm{h}) = \left(
                             \begin{array}{cc}
                               -h_1(j_1+j_2) & -2h_2j_2 \\
                               h_1j_2 & h_2j_2 \\
                             \end{array}
                           \right),
\end{equation}
and its characteristic polynomial $P(\lambda) = \lambda^2 +
a_1\lambda + a_2$, where $a_1=-h_2j_2 +h_1(j_1+j_2)$, and
$a_2=h_1h_2j_2(j_2-j_1)$. Then the stability of the stationary
states $h_1>0, h_2>0$ requires that both coefficients $a_1>0$ and
$a_2>0$ be positive simultaneously \cite{Murray}. This leads to:
\begin{equation}
j_{2} > j_{1} \;\;\;\; \mbox{and} \;\;\;\; \frac{j_1+j_2}{j_2} > \frac{h_2}{h_1}
>0
\end{equation}
as claimed.

\bibliography{bibliography_file}

\end{document}